\documentclass[fleqn,usenatbib,letters]{mnras}

\usepackage[T1]{fontenc}

\DeclareRobustCommand{\VAN}[3]{#2}
\let\VANthebibliography\thebibliography
\def\thebibliography{\DeclareRobustCommand{\VAN}[3]{##3}\VANthebibliography}

\usepackage{graphicx}	
\usepackage{amsmath}	
\usepackage{amssymb}	
\usepackage{multirow}
\usepackage{siunitx}
\usepackage{threeparttable}
\usepackage{cleveref}
\usepackage{pdflscape}
\usepackage{lscape}
\usepackage{subfigure}
\usepackage{float}
\usepackage{booktabs}

\DeclareSIUnit \parsec 	    {pc}
\DeclareSIUnit \mparsec 	{mpc}
\DeclareSIUnit \eV 			{eV}
\DeclareSIUnit \keV 		{keV}
\DeclareSIUnit \Msun 		{M_\odot}
\DeclareSIUnit \year        {yr}

\title[What does lie at the Milky Way centre?]{What does lie at the Milky Way centre? Insights from the S2 star orbit precession
}

\author[C.~R.~Arg\"uelles et al.]{
C.~R.~Arg\"uelles,$^{1,2,3}$\thanks{E-mail: carguelles@fcaglp.unlp.edu.ar (CRA)}
M.~F.~Mestre,$^{1,4}$
E.~A.~Becerra-Vergara,$^{2,4,5}$
V.~Crespi,$^{1}$ 
A.~Krut,$^{2,3}$ 
\newauthor
J.~A.~Rueda$^{2,3,6,7}$
R.~Ruffini$^{2,3,6,7}$
\\
$^{1}$Fac. de Ciencias Astron. y Geof\'isicas, Universidad Nacional de La Plata, Paseo del Bosque, B1900FWA La Plata, Argentina\\
$^{2}$ICRANet, Piazza della Repubblica 10, 65122 Pescara, Italy\\
$^{3}$ICRA, Dip. di Fisica, Sapienza Universit\`a di Roma, P.le Aldo Moro 5, 00185 Rome, Italy\\
$^{4}$Instituto de Astrofísica de La Plata, UNLP $\&$ CONICET, Paseo del Bosque, B1900FWA La Plata, Argentina\\
$^{5}$\textit{GIRG}, Escuela de F\'isica, Universidad Industrial de Santander, 680002 Bucaramanga, Colombia\\
$^{6}$ICRANet-Ferrara, Dip. di Fisica e Scienze della Terra, Universit\`a degli Studi di Ferrara, Via Saragat 1, 44122 Ferrara, Italy\\
$^{7}$INAF, Istituto de Astrofisica e Planetologia Spaziali, Via Fosso del Cavaliere 100, 00133 Rome, Italy
}

\date{Accepted XXX. Received YYY; in original form ZZZ}

\pubyear{2021}

\begin{document}
\label{firstpage}
\pagerange{\pageref{firstpage}--\pageref{lastpage}}
\maketitle

\begin{abstract}
It has been recently demonstrated that both, a classical Schwarzschild black hole (BH), and a dense concentration of self-gravitating fermionic dark matter (DM) placed at the Galaxy centre, can explain the precise astrometric data (positions and radial velocities) of the S-stars orbiting Sgr~A*. This result encompasses the 17 best resolved S-stars, and includes the test of general relativistic effects such as the  gravitational redshift in the S2-star. In addition, the DM model features another remarkable result: the dense core of fermions is the central region of a continuous density distribution of DM whose diluted halo explains the Galactic rotation curve. In this Letter, we complement the above findings by analyzing in both models the relativistic periapsis precession of the S2-star orbit. While the Schwarzschild BH scenario predicts a unique prograde precession for S2, in the DM scenario it can be either retrograde or prograde, depending on the amount of DM mass enclosed within the S2 orbit, which in turn is a function of the DM fermion mass. We show that all the current and publicly available data of S2 can not discriminate between the two models, but upcoming S2 astrometry close to next apocentre passage could potentially establish if Sgr~A* is governed by a classical BH or by a quantum DM system. 
\end{abstract}

\begin{keywords}
Galaxy: centre -- Galaxy: structure -- dark matter -- stars: kinematics and dynamics 
\end{keywords}



\section{Introduction}


The most effective method to explore the nature of the supermassive compact object at the centre of our Galaxy, Sgr A*, has been the tracing of the orbits of the S-cluster stars. This astrometric data acquisition has been performed by two leading groups, one of which started a significant progress in constraining the central object mass \citep{1997MNRAS.284..576E,2002Natur.419..694S,2009ApJ...692.1075G,2010RvMP...82.3121G,2017ApJ...837...30G}. More recently this group has incorporated the GRAVITY instrument of the VLT, allowing to detect the gravitational redshift of the S2-star \citep{2018A&A...615L..15G}; the detection of flares or hot spots close to Sgr~A* \citep{2018A&A...618L..10G}; and the detection of the relativistic precession of the S2-star orbit \citep{2020A&A...636L...5G}. In parallel a second group reached similar constraints to the mass of the central compact object in Sgr~A*   \citep{1998ApJ...509..678G,2005ApJ...620..744G,2008ApJ...689.1044G,2016ApJ...830...17B}, mainly operating with the Keck, Gemini North and Subaru Telescopes, using the Adaptative Optics technique. This group has recently confirmed the detection of the S2-star relativistic redshift \citep{2019Sci...365..664D}, converging with the first group in an estimate of the central object mass of about $4\times10^6 M_\odot$. 

From this two observational campaigns, the inference on the nature of Sgr~A* has been reached on the ground of novel theoretical understandings. A recent important result has been obtained in \citet{2020MNRAS.497.2385M}, by re-considering the flare emissions around SgrA*, emphasizing both, that their motion is not purely geodesic, and establishing a limit on the spin of a putative Kerr BH mass of $|a|<0.5$. Soon after, in \citet{2020A&A...641A..34B}, it was introduced an alternative model to the classical BH in Sgr~A* by re-interpreting it as a high concentration of quantum self-gravitating DM made of fermions of about $56$~keV/c$^2$ rest mass. This alternative approach can explain the astrometric data of both the S2-star and the G2 object with similar accuracy than the Schwarzschild BH scenario, but without introducing a drag force on G2 which is needed in the BH case to reconcile it with the G2 post-pericentre passage velocity data. An underlying assumption about the nature of such a DM quantum core is its absence of rotation, which is well supported by recent upper bounds on the spin of the central BH of $a<0.1$, based on the spatial distribution of the S-stars \citep{2020ApJ...901L..32F}. The first results obtained in \citet{2020A&A...641A..34B} within the DM scenario have been further extended in \citet{2021MNRAS.505L..64B} by considering the 17 best resolved stars orbiting Sgr~A*, achieving an equally good fit than in the BH paradigm. Remarkably, such a dense DM core is the central region of a continuous distribution of DM whose diluted halo explains the Galactic rotation curves \citep{2018PDU....21...82A,2020A&A...641A..34B, 2021MNRAS.505L..64B}. Core-halo DM distributions of this kind are obtained from the solution of the Einstein equations for a self-gravitating, finite-temperature fluid of fermions in equilibrium following the Ruffini-Arg\"uelles-Rueda (RAR) model \citep{2015MNRAS.451..622R, 2016JCAP...04..038A, 2016PhRvD..94l3004G, 2017PhRvD..96f3001G, 2018PDU....21...82A, 2019PDU....24..278A, 2020EPJC...80..183P, 2020PDU....3000699Y, 2020A&A...641A..34B, 2021MNRAS.505L..64B, 2021MNRAS.502.4227A}. These novel core-halo DM profiles, as the ones applied in this Letter, have been shown to form and remain stable in cosmological time-scales, when accounting for the quantum nature of the particles within proper relaxation mechanisms of collisionless fermions \citep{2021MNRAS.502.4227A}.

In this Letter, we focus the attention solely on the S2-star, which has been continuously monitored in the last $27$ years, and shows one of the most compact orbits around Sgr~A* with an orbital period of about $\SI{16}{\year}$ and a pericentre of $\SI{0.56}{\mparsec}$ (i.e. about $1450$ Schwarzschild radii of the $\SI{4E6}{\Msun}$ central object). Even though it is relatively far from Sgr A* where relativistic effects are feeble and hard to detect, S2 is currently considered the best tracer of the Sgr A* gravitational potential. Being its most precise astrometric data taken around pericentre passages, see e.g. \citet{2018A&A...615L..15G,2020A&A...636L...5G,2019Sci...365..664D}. 

With the aim of making progress in disentangling the nature of Sgr A*, we analyze here the relativistic periapsis precession of the S2-star in the above BH and DM scenarios. As the main result of this work, it is shown that the precession of the S2-star within the RAR DM model can be either retrograde or prograde depending upon the amount of DM mass enclosed within the orbit. The latter is shown to be a function of the mass of the DM fermion (hereafter \textit{darkino}). In particular, we show that for a $56$~keV/$c^2$ \textit{darkino} mass the RAR model predicts a retrograde S2 orbit precession, while for a slightly larger \textit{darkino} mass of $58$~keV/$c^2$, it predicts a prograde precession. The latter very much similar to the Schwarzschild BH case. By fitting all the publicly available S2 astrometric data, we conclude that none of the above scenarios about the nature of Sgr A* can be currently discriminated. This is mainly due to the large eccentricity of the S2 orbit, implying that its cumulative precession has a better chance of detectability away from the pericentre passage and closer to apocentre \citep{2017ApJ...845...22P}. Consequently, we assess at which epoch during the S2 orbital motion these scenarios can be disentangled, being the S2 high precision data beyond $2019$ of utmost importance for this task.

\section{Precession of the S2 orbit}

In both models here considered for Sgr A*, the spherically symmetric spacetime metric can be written as
\begin{equation}
        \label{eqn:metric}
        ds^2 = A(r)c^2 dt^2 - B(r)dr^2 - r^2\left(d\theta^2 + \sin^2{\theta} d\phi^2\right),
\end{equation}
where ($r$,$\theta$,$\phi$) are the spherical coordinates, $c$ the speed of light, and $A(r)$, $B(r)$ are the metric functions to be found by solving the Einstein field equations. For the Schwarzschild BH model, such equations can be solved analytically leading to: $A(r)=1 - 2 G M_{\rm{BH}}/(c^2 r)$ and $B(r)=1/A(r)$,
%
%
where $G$ is the gravitational constant and $M_{\rm BH}$ the BH mass. For the RAR DM model, the system of Einstein equations is solved numerically for $A(r)$ and $B(r)$, together with the Tolman and Klein thermodynamic equilibrium conditions, and the (particle) energy conservation along geodesics \citep{2018PDU....21...82A}. These metric potentials are not analytic and their radial dependence depend on the boundary-value problem specified to have solutions that agree with the galaxy observables. A solution of the RAR DM model for the case of the Milky Way, with specific boundary conditions that agree either with the overall rotation curve, the orbits of the 17 best resolved S-cluster stars and the G2 object, was presented in \citet{2020A&A...641A..34B, 2021MNRAS.505L..64B} for a \textit{darkino} mass of $56$~keV/$c^2$. 

A necessary condition to be fulfilled by the RAR DM profiles in order to explain the S2 orbit, is that the corresponding DM core radius ($r_c$) be smaller than the S2 pericentre (which for the case of $m=56$~keV/$c^2$ is $r_c\approx 0.4$ mpc $< r_{p(S_2)}=0.56$ mpc). However, as first understood in \citet{2018PDU....21...82A}, and further detailed here, more compact DM cores with $r_c < r_{p(S_2)}$ (with fixed Milky Way halo boundary conditions) can also explain the S-cluster stellar orbits for different \textit{darkino} masses. Thus, in this Letter we explore other Milky Way RAR profiles from $m=55$~keV/$c^2$ to $m=60$~keV/$c^2$, which will be essential to compare the properties of the S2 orbit precession for different central DM concentrations.

The equations of motion of a test particle in the spacetime metric \cref{eqn:metric}, assuming without loss of generality the motion on the plane $\theta= \pi/2$, are given by 
\begin{equation}
    \begin{aligned}
        & \dot{t} = \dfrac{E}{c^2 \ A(r)},\qquad \qquad \dot{\phi} =  \dfrac{L}{r^2},\\
        & \ddot{r} = \dfrac{1}{2 \  B(r)}\left[- A'(r) \ c^2 \ \dot{t}^2 - B'(r) \ \dot{r}^2 + 2 \ r \ \dot{\phi}^2\right],
    \end{aligned}\label{eqn:motion}
\end{equation}
%
%
where $E$ and $L$ are the conserved energy and the angular momentum of the test particle per-unit-mass, the overdot stands for derivative with respect to the proper time $\tau$, while the superscript comma ($'$) denotes derivative with respect to the radial coordinate $r$. We perform the numerical integration of eqs.~(\ref{eqn:motion}) with a Dormand-Prince algorithm  \citep{https://doi.org/10.1002/zamm.19880680638}, implemented as {\sc{DOP853}} by the Python library {\sc{SCIPY}}~\citep{2020SciPy-NMeth}. In addition, the appropriate initial conditions have been chosen in such a way that the test particle motion starts at the apocentre, i.e., $t(\tau_0) = 0$, $\phi(\tau_0)=\pi$, $r(\tau_0)=r_a$, and $\dot{r}(\tau_0) = 0$. 
We integrate the equations for a sufficiently long time which assures that the particle performs more than two consecutive orbits, so we can compute the net precession of the real orbit over two consecutive cycles. For instance, denoting the time of apocentre in two consecutive orbits respectively as $t_{\rm apo1}$ and $t_{\rm apo2}$, the precession of the real orbit over those two cycles is $\Delta \phi=\phi(t_{\rm apo2})-\phi(t_{\rm apo1})$.
\begin{figure}
  \centering
  \includegraphics[width=0.47\textwidth]{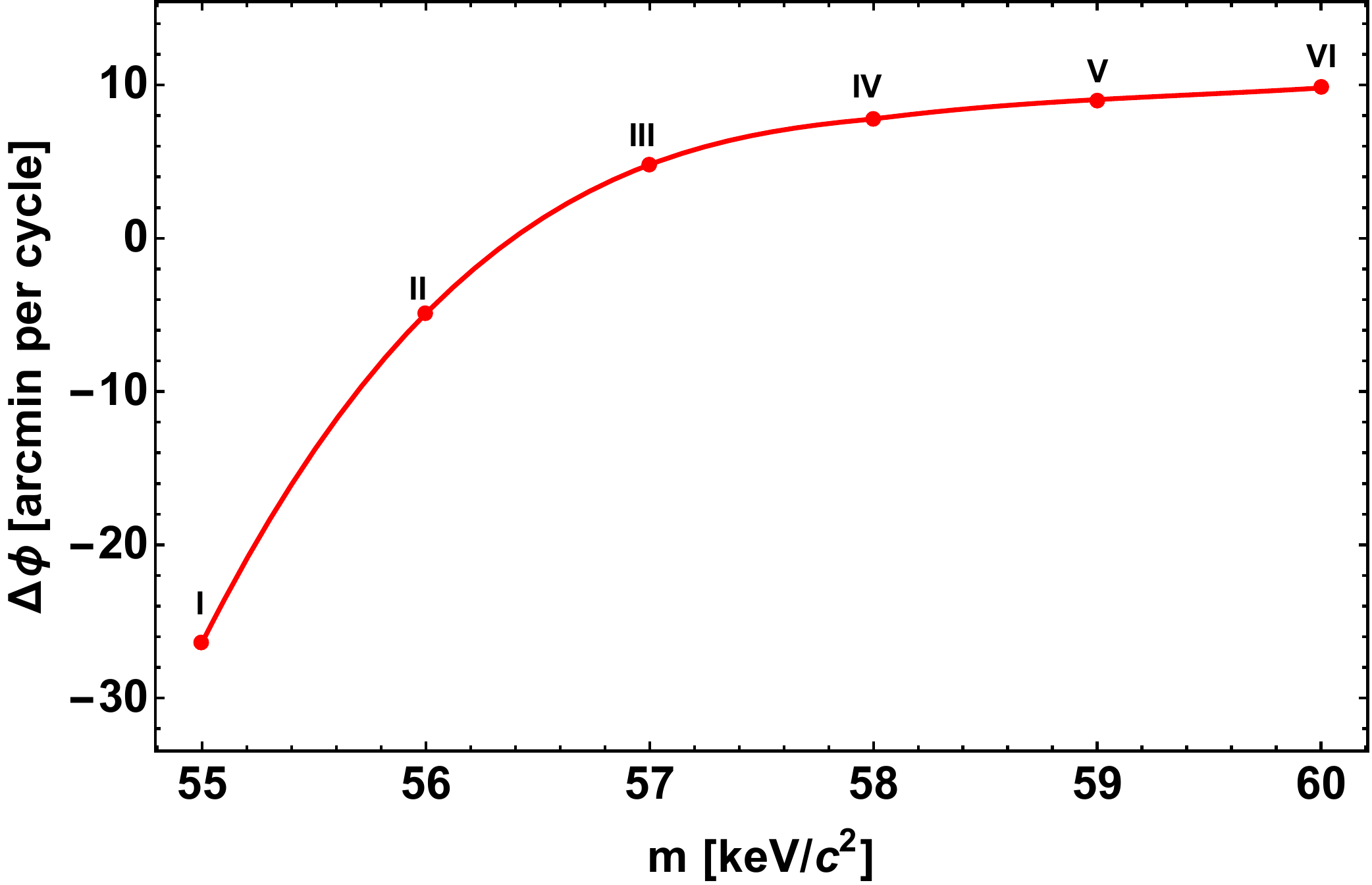}
  \caption{Relativistic periapsis precession $\Delta\phi$ per orbit as a function of the \textit{darkino} mass as predicted by the RAR DM models for the S2-star. The precession is retrograde for $m<56.4$~keV/$c^2$ while it becomes prograde for $m>56.4$~keV/$c^2$ (see  also \cref{tab:parameters}).} 
  \label{fig:precession}
\end{figure}

We start by analyzing the effects of different DM core concentrations with corresponding \textit{darkino} masses, on the precession $\Delta\phi$ (as defined above) for the S2-star. For this task, we use all the publicly available astrometric measurements \citep{2019Sci...365..664D} which include data obtained from the instruments NIRC on Keck I ($1995$--$2005$) and NIRC2 on Keck II ($2005$--$2018$). In \cref{fig:precession}, we show $\Delta\phi$ as a function of the \textit{darkino} mass $m$, within a narrow particle mass range around $56$~keV/c$^2$. In the BH case, the precession of the orbit per cycle is given by the well-known expression $\Delta\phi_{\rm BH} = 6 \pi  G M_{\rm BH}/[c^2 a (1-e^2)]$, which is always prograde ($\Delta\phi>0$) and for a $4.07\times 10^6 M_\odot$ BH mass and the S2 orbital parameters gives $\approx 12$ arcmin. In the DM case, $\Delta\phi$ has no analytic expression, but the numerical solutions show that it increases non-linearly from negative to positive (i.e. from retrograde to prograde) with the \textit{darkino} mass (see \cref{fig:precession}).

Given the DM quantum core is surrounded by an extended and more diluted DM mass, there are two competing effects in the RAR model which lead to three different possibilities for the S2 orbit precession. The effects can be roughly separated into a prograde effect caused by the gravitational potential of the DM core lying inside the S2 pericentre (similar to a relativistic point-like source), and a retrograde effect caused by the gravitational potential generated by the extended mass between the pericentre radius $r_{p}$, and the apocentre radius $r_{a}$, i.e. $\Delta M_{\rm DM} = \int_{r_p}^{r_a} 4\pi r^2\rho_{\rm DM}(r) dr$. Due to the scaling behavior of the core-halo RAR profiles with the underlying four free parameters of the theory \citep{2018PDU....21...82A,2019PDU....24..278A}, both effects are related to each other. That is, the larger the \textit{darkino} mass the more compact is the DM core (resembling more and more the BH case \citep{2018PDU....21...82A}), with consequent less extended mass fraction $\Delta M_{\rm DM}/M_c$. The above leads to: (i) Prograde precession ($\Delta\phi>0$) as shown in \cref{fig:precession} and occurring for $m>56.4$~keV/$c^2$ with a corresponding DM core mass of $M_c=3.50\times 10^6 M_\odot$. In this case, the retrograde effect due to the extended DM mass $\Delta M_{\rm DM}/M_c$ is not enough to compensate for the prograde one; (ii) Null precession ($\Delta\phi=0$) occurring for $m=56.4$~keV/$c^2$ with $M_c= 3.50\times 10^6 M_{\odot}$ when the above two effects balance each other; (iii) Retrograde precession ($\Delta\phi<0$), as shown in \cref{fig:precession} for $m<56.4$~keV/$c^2$ when the fraction of the DM core mass $\Delta M_{\rm DM}/M_c$ between $r_{p(S_2)}$ and $r_{a(S_2)}$ is large enough. As detailed in \cref{tab:parameters}, such a threshold value of $\Delta M_{\rm DM}/M_c$ below which its associated retrograde effect becomes negligible (and thus the precession is always prograde) is $\sim 0.1 \%$.

Then, we proceed to compare the predictions on the periapsis precession of the S2-orbit analogous to that of \cref{fig:precession}, i.e. calculated at apocentre, but in the plane of the sky ($\Delta\phi_{\rm sky}$) for $7$ different models: a Schwarzschild BH, and $6$ RAR DM models for the Milky Way (i.e. already reproducing its rotation curve) with $m$ from $55$ and up to $60$~keV/$c^2$. The obtained values of $\Delta\phi_{\rm sky}$ are given in last column of \cref{tab:parameters}, together with each set of free model parameters (e.g. $M_{\rm BH}$ for the BH model and $M_c$ for RAR at each given $m$) which best fits the astrometry data of S2. Some \textit{a posteriori} model properties including the DM core radius $r_c$ and the extended DM mass fraction $\Delta M_{\rm DM}/M_c$ are included in the table. The best-fit models are obtained following the procedure of \citet{2020A&A...641A..34B}, where the full set of best-fit orbital parameters of S2 (for $m=56$ keV$/c^2$) can be also found.

\begin{figure*}
  \centering
  \includegraphics[width=0.95\textwidth]{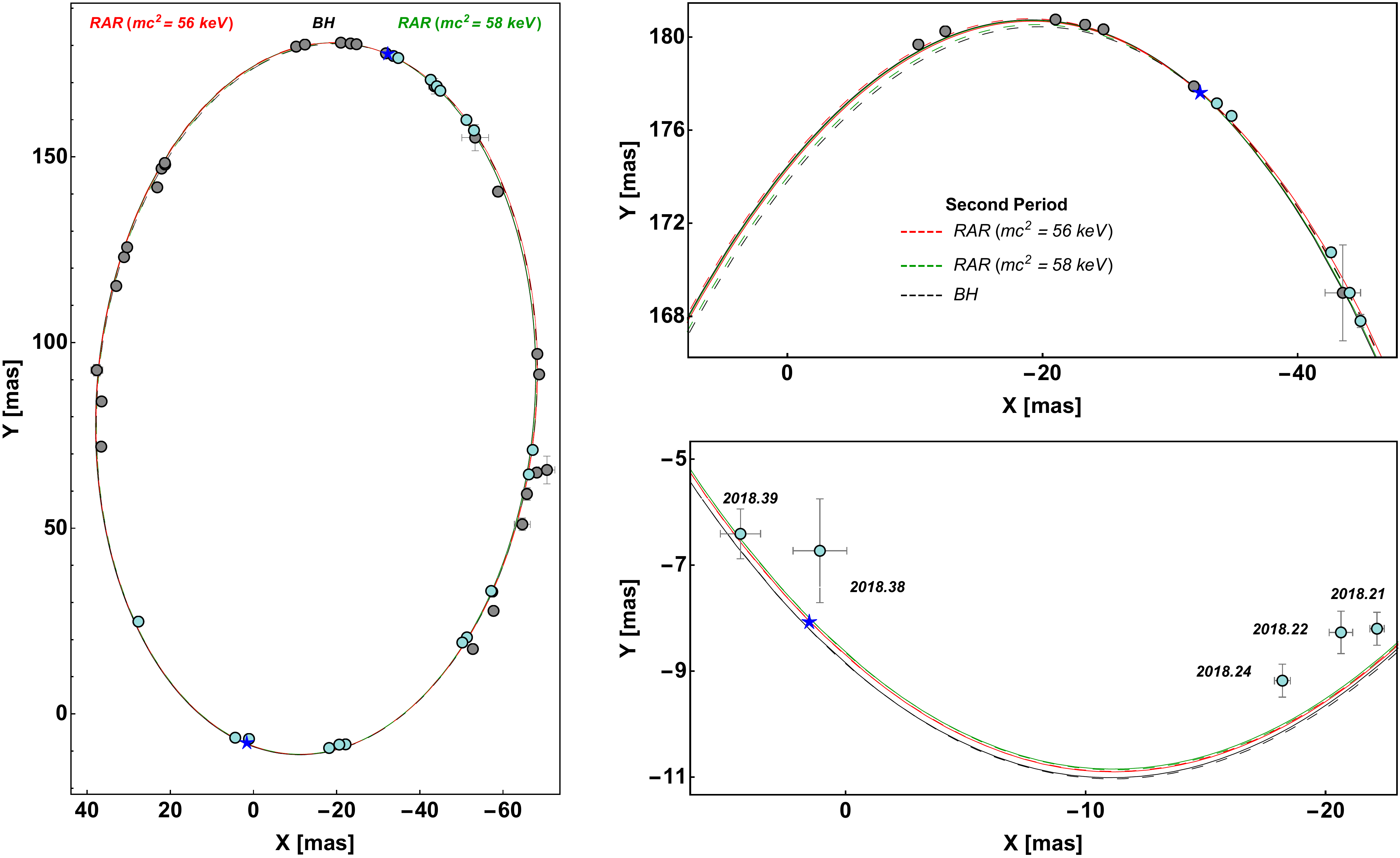}
  \caption{Relativistic precession of S2 in the projected orbit on the plane of the sky as predicted in the BH and RAR DM models. While it is prograde for the BH and RAR ($m=58$ keV$/c^2$) (in dashed black and green respectively), it is retrograde for the RAR DM model ($m=56$ keV$/c^2$) (in dashed red). The solid (theoretical) curves and gray (data) points correspond to the first period ($\approx 1994$--$2010$) while the dashed (theoretical) curves and cyan (data) points to the second period ($\approx 2010$--$2026$). \textit{Right panels}: zoom of the region around apocentre \textit{(top panel)} and pericentre \textit{(bottom panel)}. The astrometric measurements are taken from \citet{2019Sci...365..664D}.}
  \label{fig:precession_data}
\end{figure*}

\Cref{fig:precession_data} shows the relativistic precession of S2 projected orbit, in a \textit{right ascension - declination} plot. It can be there seen that while the positions in the plane of the sky nearly coincide about the last pericentre passage in the three models, they can be differentiated close to next apocentre. Specifically, the upper right panel evidences the difference at apocentre between the prograde case (as for the BH and RAR model with $m=58$~keV/$c^2$), and the retrograde case (i.e. RAR model with $m=56$~keV/$c^2$).

\begin{figure*}
   \centering
   \includegraphics[width=0.95\textwidth]{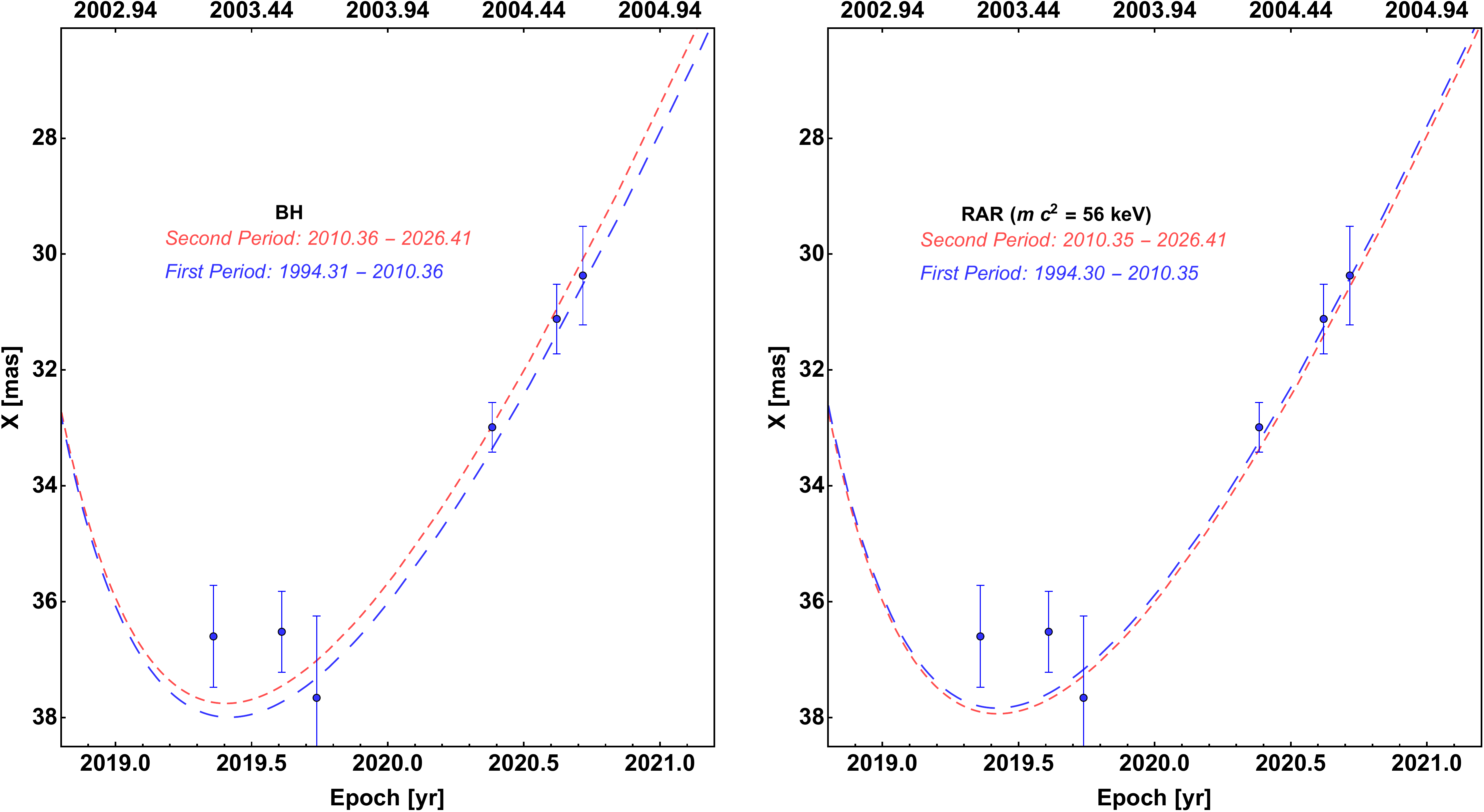}
   \caption{Relativistic precession of S2 as manifested in the right ascension as a function of time after last pericentre passage, where effects are more prominent. BH model (\textit{Left panel}) and RAR model for $m=56$ keV$/c^2$ (\textit{Right panel}).}
   \label{fig:RA_prediction}
\end{figure*}

The same conclusion can be better evidenced, and quantified, by showing the S2 orbit precession effects in right ascension (X) as a function of time, as predicted by each of the above two models. This is plotted in \cref{fig:RA_prediction} showing the clear difference in X between two consecutive periods, each one starting at about the pericentre passage. The first period is shown by the short-dashed red curve, and the second period is shown by the long-dashed blue curve which extends beyond the last  pericentre passage, where the predicted precession in each model is more evident. Indeed, the BH case (left pannel) shows a prograde trend with a maximal shift of $\Delta$X $\approx 0.7$ mas at $2026.0$ with respect to the former period (and being $\approx 0.4$ mas at $2021.2$); while the RAR ($m=56$~keV/$c^2$) case (right panel) shows a retrograde trend with a maximal shift of $\Delta$X $\approx 0.3$ mas at $2026.3$ with respect to the former period (being $\Delta$X $\approx 0.2$ mas at $2021.2$).

Unfortunately, the publicly available data in the relevant time-window of \cref{fig:RA_prediction} are only a few data points within the $1st$ period (shown in blue dots and obtained from \citet{2019Sci...365..664D}), where the large error bars impede to discriminate between the models. Even if the improved S2 astrometric resolution obtained by the GRAVITY Collaboration between $2018$ and $2019.7$ reaches the $0.1$ mas (though not yet public), it covers the range around pericentre passage where the predicted $\Delta$X in both models is too low to safely discriminate between these models.


%
%
%
%
 %
%

\begin{table*}
\centering
\caption{Comparison of the BH and RAR DM models that best fit of all the publicly available data of the S2 orbit. The $2nd$ column shows the central object mass, $M_{\rm CO}$. For the Schwarzschild BH model, $M_{\rm CO} = M_{\rm BH}$, while for the RAR model,  $M_{\rm CO} = M_c$, with $M_c$ the DM core mass. The $3rd$ column shows the radius of the central object, $r_c$. For the Schwarzschild BH model, $r_c$ is given by the event horizon radius, $R_{\rm Sch} = 2 G M_{\rm BH}/c^2$. The $4th$ column shows the DM mass enclosed within the S2 orbit, $\Delta M_{\rm DM}/M_{\rm CO}$. The best fitting pericentre and apocentre radii of the S2 orbit are given, respectively, in the $5th$ and $6th$ column. The values of the average reduced-$\chi^2$ of the best fits, defined as in \citealp{2020A&A...641A..34B}, are given in the $7th$ column. The last two columns shows, respectively, the model predictions of the periapsis precession of the real orbit, $\Delta \phi$, and of the sky-projected orbit, $\Delta\phi_{\rm sky}$.}
\label{tab:parameters}
\resizebox{\textwidth}{!}{%
\begin{tabular}{clcccccccccccccccc}
\hline
\multicolumn{2}{c}{\textbf{Model}} & & \textbf{\begin{tabular}[c]{@{}c@{}}$M_{\rm CO}$\\ $\left[10^6 M_\odot\right]$\end{tabular}} & & \textbf{\begin{tabular}[c]{@{}c@{}}$r_c$\\ $\left[\rm mpc\right]$ \end{tabular}} & & $\Delta M_{\rm DM}/M_{\rm CO}$ & & \textbf{\begin{tabular}[c]{@{}c@{}}$r_p$\\ $\left[\rm as\right]$\end{tabular}} &  & \textbf{\begin{tabular}[c]{@{}c@{}}$r_a$\\ $\left[\rm as\right]$\end{tabular}} &  &
\textbf{\begin{tabular}[c]{@{}c@{}}$\langle \bar{\chi}^2\rangle$ \end{tabular}} &  &
\textbf{\begin{tabular}[c]{@{}c@{}}$\Delta \phi$\\ $\left[\rm arcmin\right]$\end{tabular}} &  & \textbf{\begin{tabular}[c]{@{}c@{}}$\Delta\phi_{\rm sky}$\\ $\left[\rm arcmin\right]$\end{tabular}}\\
\cline{1-2} \cline{4-4} \cline{6-6} \cline{8-8} \cline{10-10} \cline{12-12} \cline{14-14} \cline{16-16} \cline{18-18} 
I   & RAR ($m = 55$ keV/$c^2$) &  & $3.55$ &  & $0.446$ &  & $1.39\times 10^{-2}$ &  & $0.01417$ &  & $0.23723$ &  & $2.9719$ &  & $-26.3845$ &  & $-32.1116$ \\
II  & RAR ($m = 56$ keV/$c^2$) &  & $3.50$ &  & $0.427$ &  & $5.99\times 10^{-3}$ &  & $0.01418$ &  & $0.23618$ &  & $3.0725$ &  & $-4.9064$ &  & $-5.9421$ \\
III & RAR ($m = 57$ keV/$c^2$) &  & $3.50$  &  & $0.407$ &  & $2.21\times 10^{-3}$ &  & $0.01417$ &  & $0.23617$ & & $3.2766$ &  & $4.8063$ &  & $5.8236$\\
IV  & RAR ($m = 58$ keV/$c^2$) &  & $3.50$  &  & $0.389$ &  & $7.13\times 10^{-4}$ &  & $0.01424$ &  & $0.23609$ & & $3.2814$ &  & $7.7800$ &  & $9.4243$\\
V   & RAR ($m = 59$ keV/$c^2$) &  & $3.50$  &  & $0.371$ &  & $2.93\times 10^{-4}$ &  & $0.01418$ &  & $0.23613$ & & $3.3356$ &  & $9.0456$ &  & $10.9613$\\
VI  & RAR ($m = 60$ keV/$c^2$) &  & $3.50$  &  & $0.355$ &  & $1.08\times 10^{-4}$ &  & $0.01423$ &  & $0.23610$ & & $3.3343$ &  & $9.8052$ &  & $11.8764$\\
      & BH &  & $4.07$ &  & $3.89 \times 10^{-4}$ &  & $0$ &  & $0.01427$ &  & $0.23623$ &  & $3.3586$ &  & $11.9501$ &  & $14.4947$ \\\hline
\end{tabular}%
}
\end{table*}
\section{Discussion and Conclusions}

In this Letter, we have demonstrated that unlike the classical Schwarzschild BH prograde precession for the S2 orbit, when assuming a quantum DM nature for Sgr~A* according to the RAR model, it can be either retrograde or prograde depending on the amount of DM mass enclosed within the S2 orbit. Such a trend, in turn, depends on the \textit{darkino} mass. We have clearly shown that within current astrometric resolution for S2, upcoming data close to the next apocentre passage have a good chance to validate one of the above predicted directions of the orbital precession. Such a confirmation is essential to unveil the classical versus the quantum nature of Sgr~A* . 

Finally, we outline some astrophysical and cosmological consequences of the RAR DM model, in addition to the present results. The RAR model predicts a mechanism for supermassive BH formation in the high-redshift Universe when the dense core of DM reaches its critical mass for gravitational collapse \citep{2021MNRAS.502.4227A}. For a \textit{darkino} mass of about $50$ keV/$c^2$, such a critical mass is $\sim 10^8 {M_\odot}$. This numerical value can be affected by additional accretion of baryonic matter on the \textit{darkino} core. Furthermore, the RAR model provides as well the quantum nature and mass of the DM particles, and the morphology of the DM profiles on inner halo scales. As recently shown in \citet{2021MNRAS.502.4227A}, the formation of \textit{core - halo} RAR DM profiles is predicted within violent relaxation mechanisms with the following key properties: (i) they form and remain stable within cosmological time-scales; (ii) they are \textit{Universal}, ranging from the scales of dwarfs up to the galaxy cluster scales \citep{2019PDU....24..278A}; and (iii) on inner halo scales, the RAR density profiles develop an extended plateau (similar to Burkert profiles), thereby not suffering from the core-cusp problem associated with the standard $\Lambda$CDM cosmology (see e.g. \citealp{2017ARA&A..55..343B}).


 %
 \section*{Acknowledgements}

C.R.A was supported by CONICET of Argentina, the ANPCyT (grant PICT-2018-03743) and ICRANet. E.A.B-V. thanks financial support from COLCIENCIAS, ICRANet--IRAP-PhD. and UIS.

\section*{Data Availability}

The astrometric data used in this work were obtained from \citet{2019Sci...365..664D}, and the data here generated are available in \Cref{tab:parameters}.
 
\bibliographystyle{mnras}

\bsp	
\label{lastpage}
\end{document}